%% file: tmm-youtube.tex
\documentclass[journal]{IEEEtran}

\usepackage{multirow}
\usepackage{url,hyperref}
\usepackage{graphicx}
\usepackage[usenames]{color}
\usepackage{amsmath}
\usepackage{flushend}

\definecolor{orange}{rgb}{1,0.5,0}

\newcommand{\secmoveup}{\vspace{-1.mm}}              
\newcommand{\bigsecmoveup}{\secmoveup\vspace{-0mm}} 
\newcommand{\textmoveup}{\vspace{-.0mm}}               
\newcommand{\itemmoveup}{\vspace{-0.mm}}               
\newcommand{\eqmoveup}{\vspace{-.0mm}}                 
\newcommand{\captionmoveup}{\eqmoveup\vspace{-2.mm}}   
\newcommand{\eat}[1]{}
\newcommand{\TODO}[1]{ {\color{blue}{\bf TODO:~{#1}}} }
\newcommand{\surl}[1]
{
	\urlstyle{same}\url{#1}
}

\begin{document}

\title{Tracking Large-Scale Video Remix\\ in Real-World Events}


\author{Lexing Xie, 
        Apostol Natsev,
        Xuming He, 
        John Kender, 
        Matthew Hill,
        and John R Smith 
\thanks{L Xie is with the Australian National University (ANU) and National ICT Australia (NICTA), an early part of this work was done while she was at IBM Research; A Natsev is with Google Research, Mountain View, CA, this work was done while he was at IBM Research; X He is with NICTA and ANU; J Kender is with Columbia University, this work is done while visiting IBM Research; M Hill and J Smith are with the IBM T J Watson Research Center.}
\thanks{NICTA is funded by the Australian Government as represented by the Department of Broadband, Communications and the Digital Economy and the Australian Research Council through the ICT Centre of Excellence program. 
This work is partially supported by the US Air Force Research Laboratory, under agreement number FA2386-12-1-4041. 
The views and conclusions contained herein are those of the authors and should not be interpreted as necessarily representing the official policies or endorsements, either expressed or implied, of the Air Force Research Laboratory or the U.S. Government.
}

} 

\maketitle

\secmoveup
\begin{abstract}
Social information networks, such as YouTube, contains traces of both explicit
online interaction (such as ``like'', leaving a comment, or subscribing to video feed),
and latent interactions (such as quoting, or remixing parts of a video).
We propose visual memes, or frequently re-posted short video segments, for detecting and monitoring such latent video interactions at scale. Visual memes are extracted by scalable detection algorithms that we develop, with high accuracy. We further augment visual memes with text, via a statistical model of latent topics.  We model content interactions on YouTube with visual memes, defining several measures of influence and building predictive models for meme popularity. Experiments are carried out on with over 2 million video shots from more than 40,000 videos on two prominent news events in 2009: the election in Iran and the swine flu epidemic. In these two events, a high percentage of videos contain remixed content, and it is apparent that traditional news media and citizen journalists have different roles in disseminating remixed content. We perform two quantitative evaluations for annotating visual memes and predicting their popularity. The joint statistical model of visual memes and words outperform a concurrence model, and the average error is is $\pm 2\%$ for predicting meme volume and $\pm 17\%$ for their lifespan. 
\end{abstract}

\IEEEpeerreviewmaketitle

\input{intro}

\input{meme-youtube}

\input{meme-detection}

\input{meme-topics}

\input{meme-graph}

\input{experiments}

\input{related}

\input{conclusion}
\secmoveup
\bibliographystyle{IEEEtranS}
\bibliography{youtube}


\end{document}

%% file: intro.tex

\bigsecmoveup
\section{Introduction}
\textmoveup


The ease of publishing and sharing videos~\cite{ytstat} has outpaced the progress of
modern search engines, collaborative tagging sites, and content aggregation
services---leaving users to see only a fraction of a subject.
This information overload problem is particularly
prominent for linear media (such as audio, video, animations), where
at-a-glance impressions are hard to develop and are often unreliable.  
While text-based information networks such as Twitter 
rely on retweets~\cite{Kwak10twitter}, hashtags~\cite{Romero2011hashtag} and mentions 
to identify influential and trending topics, 
similar functions for large video-sharing repository are lacking. 
On the other hand, video clipping or remixing is an essential part of the 
participatory culture on sites like YouTube~\cite{snickars10youtube}. 
Moreover, a reliable video-based ``quote'' tracking and popularity analysis system
would find utility in many different domains,
such as brand monitoring, event spotting for 
emergency management, trend prediction, journalistic content selection, 
better retrieval, or improved social data sampling systems.

We propose visual meme, 
a short segment of video that
is frequently remixed and reposted by more than one author,
as a tool for making sense of video ``buzz''.  
Video-making requires significant effort and time,
therefore reposting a video
meme is a deeper stamp of approval or awareness than simply leaving a
comment, giving a rating, or sending a tweet.
Example video memes are shown in Figures~\ref{fig:dupexample}, 
represented in a static keyframe format. 
We develop a large-scale
event monitoring system for video content, 
using generic text queries as a pre-filter
for content collection on a given topic. 
We deploy this system for YouTube, and collect large video 
datasets over a range of topics. 
We then perform fast and accurate visual meme 
detection on tens of thousands of videos and millions of video shots. 
We augment visual memes with text using statistical topic models;
we then propose a Cross-Modal Matching (CM$^2$) method that explains 
a visual meme with keywords.
We design a graph representation for social interactions via visual memes,
and then derive graph metrics 
to capture content influence and user roles.  
Furthermore, we use features derived from the video content and meme interactions to 
predict popular memes, with an average error within 2\% on the volume and 17\% of the lifespan. 

This work is an expanded version of a conference publication on visual memes~\cite{xie2011visualmeme}, another study on video duplication patterns is based on the same data collection and video remix detection method~\cite{Kender2010VGC}. The following components are new since~\cite{xie2011visualmeme}: CM$^2$, a new representation for visual memes and their associated text, new approach and results on predicting meme popularity, and updated and expanded discussions on related work. The main contributions of this work are as follows:
\begin{itemize}
\itemmoveup \item
We propose {\em visual memes} as a novel tool to track 
large-scale video remixing in social media. 
We implement a scalable system that extracts all
memes from over a million video shots, in a few hours on a single desktop computer.  
\itemmoveup \item
We design and implement the first large-scale event-based social video monitoring and content analysis system.
\itemmoveup \item
We design a novel method, CM$^2$, to explain visual memes with statistical topic models.
\itemmoveup \item
We design a graph model for social interaction via visual memes for characterizing information 
flow and user roles.
\itemmoveup \item
We conduct empirical analysis on several large-scale event datasets, 
producing observations about the extent of video remix,
the popularity of memes against traditional metrics,
and different diffusion patterns.
\end{itemize}
 
The rest of this paper is organized as follows. Section~\ref{sec:vdup} presents a system for event monitoring on YouTube; Section~\ref{sec:neardup} covers the meme detection algorithm; Section~\ref{sec:topic} proposes CM$^2$ model for annotating visual memes; Section~\ref{sec:graph} explores graph representation for meme diffusion and Section~\ref{sec:gfeatures} uses graph features to predict content importance; Section~\ref{sec:exp} discusses experiment setup and results; Section~\ref{sec:related} reviews related work; Section~\ref{sec:conclusion} concludes this work.

%% file: meme-youtube.tex

\secmoveup
\section{Visual remix and event monitoring}
\label{sec:vdup}
\textmoveup

In this section, we define visual memes as the unit for video remix tracking, and describe our system to monitor video traces from real-world events.

\begin{figure*}[tb]
\begin{minipage}[b]{0.53\linewidth}
\centering \includegraphics[angle=0,width=\textwidth]{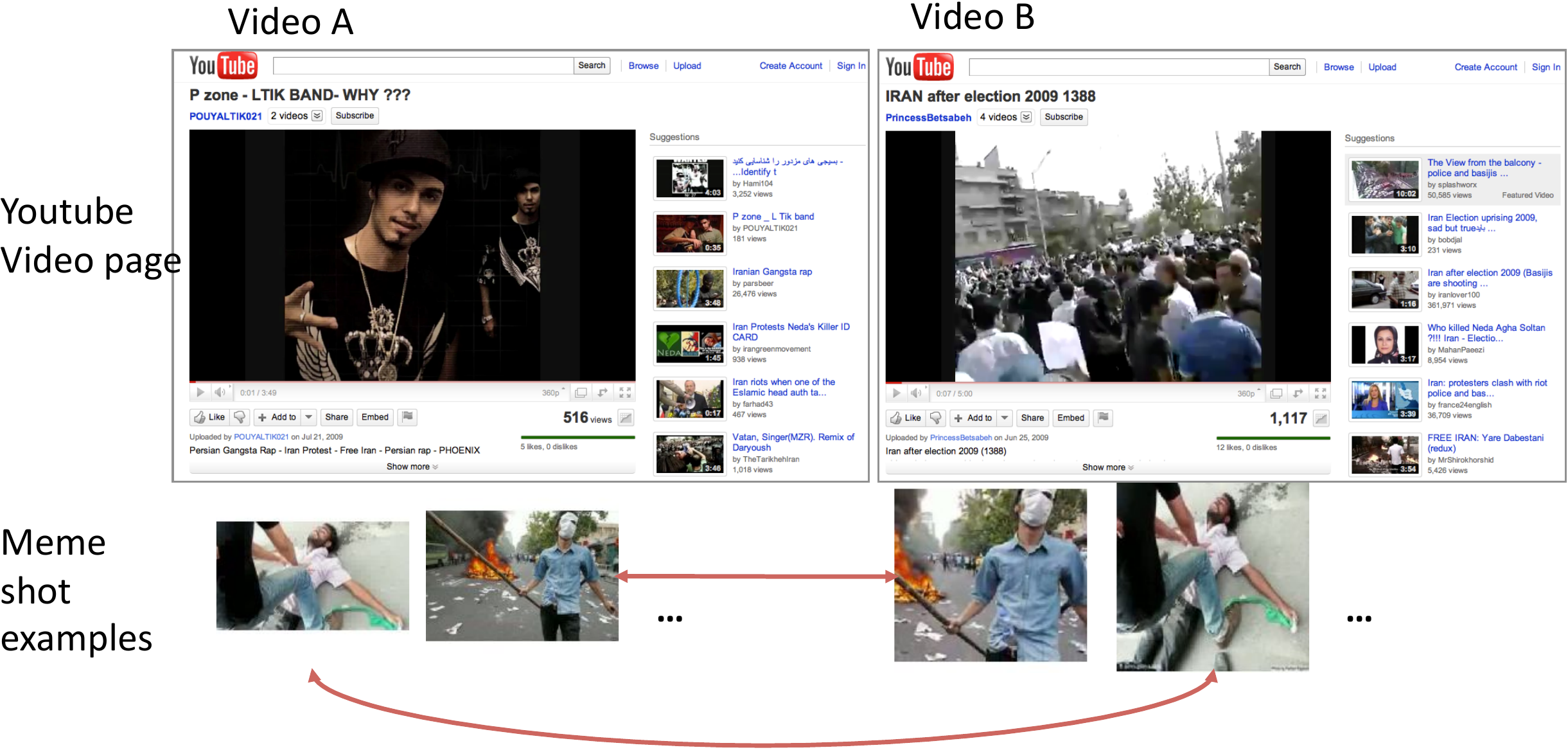}
\end{minipage}
\hspace{2mm}
\begin{minipage}[b]{0.4\linewidth}
\centering \includegraphics[angle=0,width=\textwidth,height=48mm]{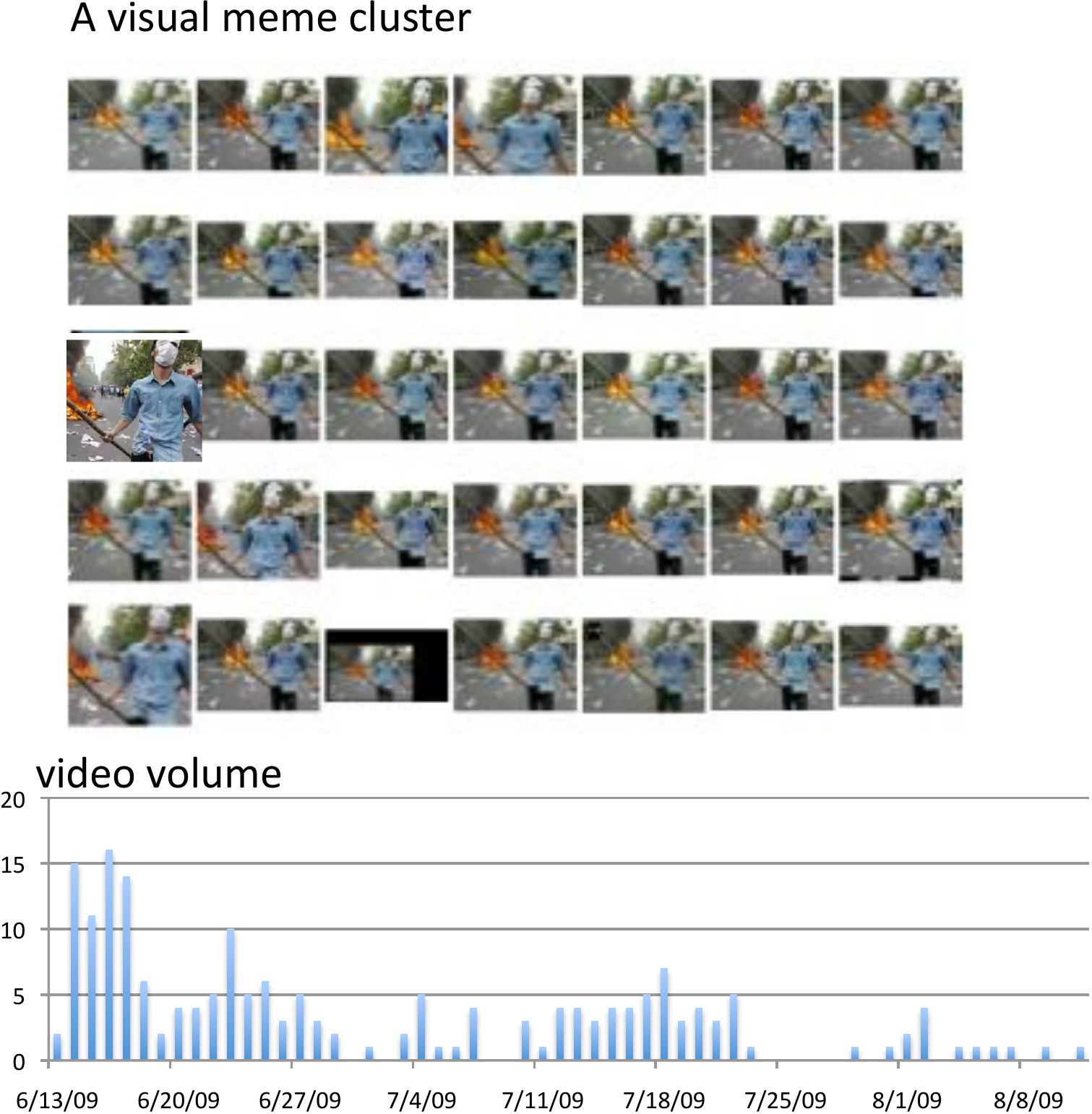}
\end{minipage}
\caption{
Visual meme shots and meme clusters. 
(Left) Two YouTube videos that share multiple different memes. 
Note that it is impossible to tell from metadata or the YouTube video page 
that they shared content, and
that the appearance of the remixed shots (bottom row) has large variations.
(Right) A sample of other meme keyframes corresponding to one of the meme shots, and the number of videos containing this meme over time -- 193 videos in total between June 13 and August 11, 2009.}
\label{fig:dupexample}\captionmoveup
\end{figure*}

\secmoveup
\subsection{Visual memes and online participatory culture}
\textmoveup

\eat{
\TODO{R2: The paper would be stronger if it first presented a solid discussion of what a meme actually is, first independently of the YouTube context and then within the YouTube context. Then, it should note the connection between near-duplicate clips, user-contributed text and memes. The limitations of looking at memes in this way should be clearly stated up front.}
}
The word {\em meme} originally means~\cite{oxdict_meme}
``an element of a culture or system of behaviour passed from one individual to another by imitation or other non-genetic means'', and on digital artifact, it refers to
``an image, video, piece of text, etc., typically humorous in nature, that is copied and spread rapidly by Internet users, often with slight variations''.
The problem of automatically tracking online memes has been recently solved for text quotes~\cite{leskovec2009meme} from news and blogs, as well as for edited images~\cite{Kennedy08imagecopy} from web search. 
The scope of this work is to study memes in 
online video sharing, and in particular, on YouTube. 

Media researchers observe that users tend to create 
``curated selections based on what they liked
or thought was important''~\cite{snickars10youtube}, 
and that remixing (or re-posting video segments)
is an important part of the ``participatory culture''~\cite{burgess2009youtube} of YouTube. 
Intuitively, re-posting 
is a stronger endorsement requiring much more effort 
than simply viewing, commenting on, or linking to the video content. 
A re-posted visual meme is an
explicit statement of mutual awareness, or a relevance statement on a
subject of mutual interest.  Hence, memes can be used to study virality, lifetimes and
timeliness, influential originators, and (in)equality of reference.

We define visual memes as frequently reposted video segments or images, 
and this study on video remix has two operational assumptions.
The first is to focus on videos about particular news events. 
Using existing footage is common practice in the 
reporting and discussion of news events~\cite{machin2006archive}.
The unit of reuse typically consists of one or a few contiguous shots, 
and the audio track often consist of re-dubbed commentary or music. 
The second assumption is to restrict remix-tracking 
to short visual segments. 
Thus we will not detect image sequence retargeted 
for entirely different meanings using redubbed audio track.  
The remixed shots typically contain minor modifications 
that include video formatting changes (such as
aspect ratio, color, contrast, gamma) and production edits (such as
the superimposing text, or adding borders and transition effects).
Most of such transformations are well-known 
as the targets of visual copy detection benchmark~\cite{natsev2010design}.

Using this definition and its assumptions, 
this work provides tools for detecting visual remixes, 
builds prototype systems that can quantify the 
prevalence of remixes. 
The observations specific to news topics do not readily generalize to 
the entire YouTube, or to video genres designed for creativity 
and self-expression, such as video blogs.
In the rest of this paper, {\em meme} refers both to 
individual instances, visualized as representative
icons (as in Figure~\ref{fig:dupexample} Left), 
and to the entire equivalence class of re-posted near-duplicate video segments,
visualized as clusters of keyframes 
(as in Figure~\ref{fig:dupexample} Right).

\secmoveup
\subsection{Monitoring Events on YouTube}
\label{sec:ytmonitor}
\textmoveup


We use text queries to pre-filter content, and make the scale of monitoring feasible~\cite{ytstat}. 
We use a number of generic, time-insensitive text queries as content pre-filters. 
The queries are manually
designed to capture the topic theme, as well as the generally understood cause,
phenomena, and consequences of the topic. For example, our queries 
about the ``swine flu'' epidemic
consists of {\em swine flu, H1N1, H1N1 travel advisory, 
swine flu vaccination}\footnote{The full set of queries is available on the accompanying webpage~\cite{projpage}}. We aim to create queries covering the
main invariant aspects of a topic, but automatic time-varying query
expansion is open for future work. We use the YouTube
API to extract video entries for each query, sorted by relevance and recency, respectively.
The API will return up to 1000 entries per query, so
varying the ranking criteria helps to increase content coverage and diversity. 
Then, for each unique video, we segment it into shots 
using thresholded color histogram differences.
For each shot we randomly select and extract a frame as keyframe, and extract visual
features from each keyframe.  We process the metadata associated with each video, 
and extract information such as author, publish date, view counts, and
free-text title and descriptions. We clean the free-text metadata
using stop word removal and morphological normalization.
The volume of retrieved and memes are telling indicators of event evolution
in the real world, a few example trends can be found in our recent paper~\cite{xie2011visualmeme} and project webpage~\cite{projpage}.

%% file: meme-detection.tex
\secmoveup
\section{Scalable Visual Meme Detection}
\label{sec:neardup}
\textmoveup
Detecting visual memes in a large video collection is a non-trivial problem. 
There are two main challenges.  First,
remixing online video segments changes their visual appearance, adding noise 
as the video is edited and re-compressed (Section~\ref{sec:vdup}). 
Second, finding all pairs of near-duplicates by matching
all N shots against each other has a complexity of $O(N^2)$,
which is infeasible for any collection containing more
than one million video shots.

Our solution to keyframe matching has three parts, 
each contributing to the robustness of the match.
Here a keyframe is representative of a video shot, 
segmented using temporal feature differences.
We first pre-process the frame by removing trivial (e.g. blank) matches, 
detecting and removing internal
borders; normalizing the aspect ratio; de-noising with median filters; 
and applying contrast-limited histogram equalization to correct for contrast and
gamma differences.
We then extract the
{\em color correlogram}~\cite{huang99correlogram} feature for each frame to capture
the local spatial correlation of pairs of colors.  The color
correlogram is designed to tolerate moderate changes
in appearance and shape that are largely color-preserving, e.g.,
viewpoint changes, camera zoom, noise, compression, and to a smaller
degree, shifts, crops, and aspect ratio changes.
We also use a ``cross''-layout that extracts the
descriptor only from horizontal and vertical central image stripes,
thereby emphasizing the center portion of the image and 
improves robustness with respect to
text and logo overlay, borders, crops, and shifts.
We extract an auto correlogram in a
166-dimensional perceptually quantized HSV color space, resulting in a
332-dimensional feature. 
Finally, the system uses a query-adaptive threshold to normalize among 
the query frame, and among the different feature dimensions. 
This threshold is tuned on a training set. 
Detailed comparison of each technique can be found in a related paper~\cite{natsev2010design}.

Our solution to the complexity challenge is to use an indexing scheme for
fast approximate nearest neighbor (ANN) look-up.  We use the FLANN Library
~\cite{flann}
to automatically select the best indexing structure 
and its appropriate parameters for a given dataset. 
\eat{FLANN allows us to set the maximum number of candidate nodes $m$ to check 
in a search, so when each query runtime is bound to $O(m)$, 
the running of $N$ queries against the entire set of $N$ keyframes can be
accomplished in $O(Nm)$ time.  }
Our frame features have over 300 dimensions, and we empirically found that 
setting the number of nearest-neighbor candidate nodes $m$ to 
$\sqrt{N}$ can approximate $k$-NN results with approximately 
$0.95$ precision.  In running in $O(N\sqrt{N})$ time, it
achieves two to three decimal orders of magnitude speed-up over exact nearest neighbor search.
Furthermore, each FLANN query results in an incomplete set of 
near-duplicate pairs, we
perform transitive closure on the neighbor relationship 
to find equivalence classes of near-duplicate sets.  
We use an efficient set union-find
algorithm that runs in amortized time of
$O(E)$,
where $E$ is the number of matched pairs~\cite{union-find},
which is again $O(N\sqrt{N})$. 

\begin{figure}[tb]
\centering \includegraphics[angle=0,width=.48\textwidth]{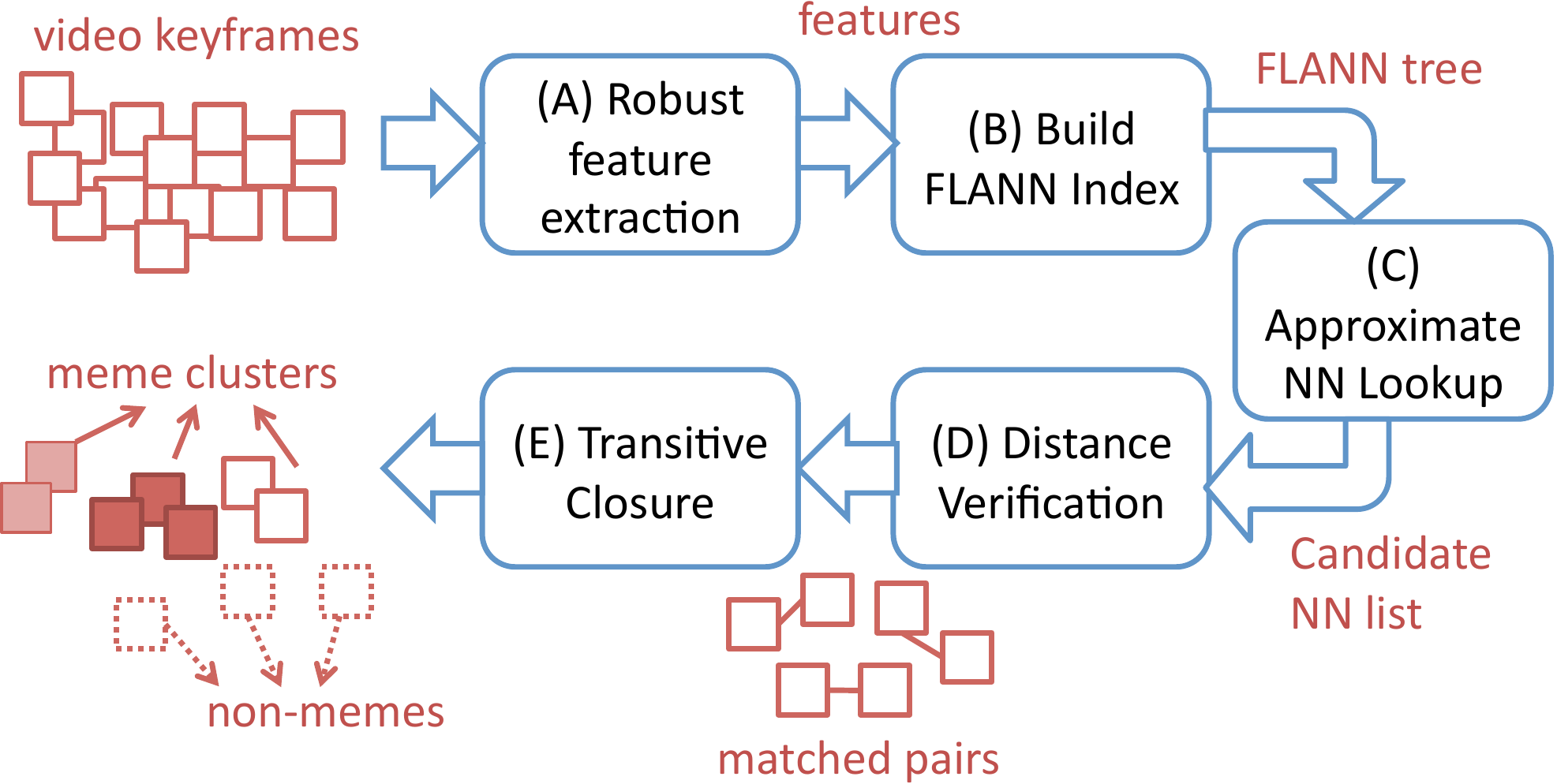} 
\captionmoveup
\caption{Flow diagram for visual meme detection method.}
\label{fig:detflow} \captionmoveup
\end{figure}

This process for detecting video memes is outlined in Figure~\ref{fig:detflow}. 
The input to this system is a set of video frames, and the output splits this 
set into two parts. The first part consists of a number of meme clusters, 
where frames in the same cluster are considered near-duplicates with each other.
The second part consists of the rest of the frames that are not considered near-duplicates
with any other frame. Blocks A and D address the robust matching challenge 
using correlogram features and query-adaptive thresholding, 
and blocks B, C and E address the scalability challenge 
using approximate nearest-neighbor (ANN) indexing. 
A few examples of identified near-duplicate sets are shown in
Figure~\ref{fig:dupexample}.  Visual meme detection performance 
is evaluated in Section~\ref{ssec:neardups}.

Our design choices for the visual meme detection system 
aim to find a favorable combination of accuracy and 
speed feasible to implement in one single PC. 
Note that local image points or video sequences~\cite{tan09accelerating} tend to be
accurate in each query, but is not easy to scale to $N^2$ matches. 
We found that a single video shot is a suitable unit to capture community video remixing,
and that matching by video keyframe is amenable to building fast indexing structures. 
The ANN indexing scheme we adopt scales to several million video shots. 
On collections about tens of millions to billions video shots,
we expect that the computation infrastructure will need to change, such as 
using a data center to implement a massively distributed tree 
structure~\cite{liu07clustering} and/or hybrid tree-hashing techniques.

%% file: meme-topics.tex

\secmoveup
\section{Topic representation of memes}
\label{sec:topic}
\textmoveup

Visual memes are iconic representations for an event, 
it will be desirable to augment its image-only representation with textual explanations. 
It is easy to see that the title and text descriptions associated 
with many online videos can be used for this purpose, despite the noisy 
and imprecise nature of the textual content. 
We propose to build a latent topic model over both the
visual memes and available text descriptions, in order to 
derive a concise representation of videos using memes, 
and to facilitate applications such as annotation and retrieval.


Our model treats each video as a document, in which visual memes are ``visual words'' 
and the annotations are text words. By building a latent topic space for video
document collections, we embed the high-dimensional bag-of-words into a more
concise and semantically meaningful topic space. 
Specifically, we learn a set of topics $z=1,\ldots,K$ on the
multimedia document collection ${\cal D} = \{d_m, m=1,\ldots,M\}$ using 
latent Dirichlet Allocation (LDA)~\cite{blei2003latent}. LDA models each document 
as a mixture of topics with a document-dependent Dirichlet prior, each
topic drawn from the resulting multi-nomial, and each word drawn from
a topic-dependent multi-nomial distribution. Our LDA model combines two types of words,
i.e., visual memes and text words, into a single vocabulary $\mathcal{V}=\{\mathcal{V}_v, \mathcal{V}_t\}$,
and estimates the conditional distribution $\Phi$ of words given topics from a video collection.
Mathematically, each video $d_m$ is represented as a bag of words, $\mathbf{w}_m=\{w_i\}_{i=1}^{N_m}$, is modeled as follows,
\begin{align}
 P(d_m|\alpha) = \int_\theta \prod_{w_i\in d_m} [\sum_{z_i} P(w_i|z_i, \Phi)P(z_i|\theta)] P(\theta|\alpha) d\theta 
\end{align}
where $z_i$ is the topic indicator variable for word $w_i$, $\theta$ is the latent topic distribution, and 
$\alpha$ is the hyperparameter controlling the topic prior at the corpus level. 

Given the topic model, we can project a set of visual words (or text tags) into the learned topic space by computing 
the posterior of the topic distribution $\theta$ conditioned on those words. Let the observed words are $\mathbf{w}_o$,
we map $\mathbf{w}_o$ to the mode $\hat \theta$ of the topic posterior:
\begin{align}
 \hat \theta = \arg\max_\theta \prod_{w_i\in\mathbf{w}_o}\sum_{z_i}P(w_i|z_i)P(z_i|\theta,\Phi) P(\theta|\alpha)\label{eq:topics}
\end{align}
where the parameters $\alpha$, $\Phi$ are estimated from training data. The inference in model learning and posterior 
calculation are conducted with variational EM (e.g., see~\cite{blei2003latent} for details).

\eat{We obtain the following
quantities from the LDA model: (1) The conditional distribution of
word given topic $\Phi$ for each element, $\Phi_{k,i}=p(w=v_i|z=k)$.
(2) The Dirichlet parameters of topics given a document $\alpha$,
which can be normalized to give expections of topic prior for a
document, $\hat \theta_{m,k}=p(z= k|d_m)$.}  

\secmoveup 
\subsection{Cross-modal matching (CM$^2$) with topics} 
\label{sec:cm2}\textmoveup

In social media, some words and names may be unseen in previous events (e.g. {\em entekhabat}, ``election'' in Persian), and iconic visual memes
may appear without clear context of emergence. For a better understanding of these novel events, a particular useful step is to build association between different modalities, 
such as texts and visual memes. 
We pose this as a cross-modal matching problem, and aim to estimate how well a
textual or visual word (candidate result $w_r$) can explain another set of words (query $\mathbf{w}_q=\{w_{q_n}\}$). This is achieved by estimating the
conditional probability of seeing $w_r$ given that $\mathbf{w}_q$ is in the document, i.e., $p(w_r|\mathbf{w}_q,D)$. We
call this estimation process \underline{C}ross-\underline{M}odal \underline{M}atching (CM$^2$), and propose its application for
content annotation and retrieval.

A derivation sketch for CM$^2$ is as follows, under the context of document collection $D$ and the topic model $\{\alpha, \Phi\}$. We
consider modeling the conditional word distribution through topic representation. Without loss of generality, we assume the query consists of visual memes and predict 
the probability of each text tag.  The first step of our method is to compute the topic posteriors of the 
document collection $D=\{d_m\}$ given the query modality. Let $\mathbf{w}_o$ be the observed visual memes in each document $d_m$, we estimate 
the topic posterior mode $\hat \theta_m$ from Equation~\ref{eq:topics}. Thus the whole document collection can be represented as $\Theta=\{\theta_m\}$. 

Given a new query $\mathbf{w}_q$, we also embed it into the topic space by computing its topic posterior mode:
\begin{align}
 \theta_q = \arg\max_\theta \prod_{w_i\in\mathbf{w}_q}\sum_{z_i}P(w_i|z_i)P(z_i\theta,\Phi) P(\theta|\alpha)\nonumber
\end{align}
Intuitively, we want to use ``similar'' videos in the topic space to predict the text tag probability. Formally, the conditional probability of a text tag $w_r$ 
is estimated by a non-parametric voting scheme as follows,
\begin{align}
 P(w_r|\mathbf{w}_q,D)\propto \sum_m \left(\sum_{w_i\in d_m} [w_i=w_r]\right)e^{-(\theta_q-\theta_m)^2/\sigma_\theta}\label{eq:topic_tag}
\end{align}
where $\sigma_\theta$ controls the similarity of topic vectors and is set to the median of the training data.

A baseline method based on word co-occurrence can estimate the conditional probability with co-ocurrence counting:
\begin{align}
 P(w_r|\mathbf{w}_q,D)\propto \sum_m \left(\sum_{w_i\in d_m,q_n} [w_i=w_r]\wedge [w_{q_n}\in\mathbf{w}_q]\right)\label{eq:word_tag}
\end{align}

Examining the estimation equations~(\ref{eq:topic_tag})-(\ref{eq:word_tag}), we note that CM$^2$ can be interpreted as a soft
co-occurrence measure for $(w_r,\mathbf{w}_q)$ over the entire document collection with the topic model. In a sense, co-occurrence counting 
is a special case, where the counts are weighted by the number of documents 
in which $\mathbf{w}_q$ appeared.

\subsection{CM$^2$ applications}

CM$^2$ has several applications depending on the choice of $w_q$ and $w_r$. Such as \eat{(1) Visual meme annotation -- Finding words that
best describe a visual meme. Here we take $w_q\in {\cal V}_v$, and return the top entries of $w_r\in {\cal V}_t$, sorted by
$p(w_r|w_q,D)$.} (1) Visual Meme/video annotation -- We use visual memes as queries, $\mathbf{w}_q\subset {\cal V}_v$, and return the top entries
of $w_r\in {\cal V}_t$, sorted by $p(w_r|\mathbf{w}_q,D)$. The motivation of this task for event monitoring is that the keywords are
often specialized subjective, semantic, and non-visual, e.g {\em freedom}.\eat{, or {\em keshavarz} (a Persian scholar). Illustrating memes
in the context of other semantically related memes helps identify the common context.} (2) Keyword illustration -- We can illustrate a
keyword (e.g.~{\em H1N1}) with a set of most-related images. We take $w_q\in {\cal V}_t$, and yield the top entries of $w_r\in {\cal V}_v$,
sorted by $p(w_r|w_q,D)$. In this paper, we focus on application (1) and leave the others for future exploration.

%% file: meme-graph.tex
\secmoveup
\section{Graphs on Visual Memes}
\label{sec:graph}
\textmoveup

Visual memes can be seen as implicit {\em links} between videos and their creators
that share the same unit of visual expression. We construct graph representations 
for visual memes and users who create them. This gives us a novel way to quantify influence 
and the importance of content and users in this video-centric
information network.

Denote a video (or multimedia document) as $d_m$ in event collection
${\cal D}$, with $m=1,\ldots,N$. Each video is authored (i.e., uploaded) by 
author $a(d_m)$ at time $t(d_m)$, with $a(d_m)$ taking its value from
a set of authors ${\cal A}=\{a_r, r=1,\ldots,R\}$. Each video document $d_m$
contains a collection of memes, $\{v_{1}, v_{2},\ldots,v_{K_m}\}$ 
from a meme dictionary ${\cal V}$. 
In this network model, each meme induces a
time-sensitive edge $e_{mj}$ with creation time
$t(e_{mj})$, where $m,j$ are over video documents or authors.

\secmoveup
\subsection{Meme video graph}
\textmoveup

We define the video graph $G=\{{\cal D}, {\cal E}_G\}$, with 
nodes $d \in {\cal D}$. There is a directed edge $e_{mj} \in {\cal E}_G$, if
documents $d_m$ and $d_j$ share at least one visual meme, and if $d_m$
precedes $d_j$ in time, with $t(d_m) < t(d_j)$. 
The presence of $e_{mj}$ represents a probability
that $d_j$ was derived from $d_m$, even though there is no
conclusive evidence within the video collection alone whether or not this
is true. We denote the number of shared visual memes as
$\nu_{mj}=|\,d_m \cap d_j\,|$, and the time elapsed between the posting time
of the two videos as $\triangle t_{jm} = t(d_j) - t(d_m)$.

We use two recipes for computing the edge weight $\omega_{mj}$.
Equation~\ref{eq:diffushareddup} uses a weight proportional to
the number of common memes $\nu_{mj}$, and Equation~\ref{eq:diffuweighted} 
scales this weight by a power-law memory factor related to the time difference $\triangle t_{jm}$. 
The first model is insensitive to $\triangle t_{jm}$, so it can accommodate 
the resurgence of popular memes, 
as seen in textual memes~\cite{leskovec2009meme}. 
The power law decay comes from known behaviors on
YouTube~\cite{crane2008robust}, and it also agrees with our observations on the recency of 
the content returned by the YouTube search API.
\begin{eqnarray}
\eqmoveup
\omega^*_{mj} =& \nu_{mj}  & (m,j) \in {\cal E}_G
\label{eq:diffushareddup}\\
\omega'_{mj} =& \nu_{mj} \triangle t_{jm}^ {-\eta} 
\label{eq:diffuweighted}
\eqmoveup
\end{eqnarray}
We estimate the exponent $\eta$ to be $0.7654$, by fitting an exponential curve to the video age versus volume to a subset of our data,
over ten different topics retrieved over 24 hours of time. 

\secmoveup
\subsection{Meme author graph}
\textmoveup

We define an author graph
$H=\{{\cal A}, {\cal E}_H\}$, with 
each author $a \in {\cal A}$ as nodes. 
There is an undirected edge $e_{rs} \in {\cal E}_H$, if
authors $a_r$ and $a_j$ share at least one visual meme
in any video that they upload in the event collection.

We 
compute the edge weights $\theta_{rs}$ on edge $e_{rs}$ 
as the aggregation of the weights on all the 
edges in the video graph $G$ connecting
documents authored by $a_r$ and $a_s$.
\begin{eqnarray}
\theta_{rs} &=& \Sigma_{\{i, a(d_m)=a_r\}} \Sigma_{\{j, a(d_j)=a_s\}}
\omega_{mj} \label{eq:agraphweight}
\end{eqnarray}
with $r,s \in {\cal A},~m,j \in {\cal D}$. 
We adopt two simplifying assumptions in this definition.
The set of edges ${\cal E}_H$ are bidirectional,
as authors often repost memes from each other at different times. 
The edge weights are cumulative over time,
because in our datasets most authors post no more than a handful of
videos (Figure~\ref{fig:diffusion}), and there is rarely enough data to estimate
instantaneous activities.

\secmoveup
\subsection{Meme influence indices}
\label{ssec:diffuseidx}
\textmoveup

We define three indices based on meme graphs, which captures 
the influence on information diffusion among memes, and in turn quantifies the
impact of content and of authors within the video sharing information network. 

First, for each visual meme $v$, we extract from the event collection
${\cal D}$ the subcollection containing all videos that have
at least one shot matching meme $v$, denoted as ${\cal D}_v=\{d_j \in {\cal
D}, \mbox{s.t.} ~v \in d_j\}$.  We use ${\cal D}_v$ to extract the
corresponding video document subgraph $G_v$ and its edges, setting
all edge weights $\nu$ in $G_v$ to $1$ since only a single meme is involved. 
We compute the in-degree and out-degree of every video $d_m$ in
${\cal D}_v$ as the number of videos preceding and following $d_m$ in time:\\
\eqmoveup
\begin{eqnarray}
\zeta^{in}_{m, v} = \Sigma_{j} I\{d_m, d_j \in {\cal D}_v, & t(d_j)<t(d_m) \} \label{eq:inoutdeg}\\
\zeta^{out}_{m, v} = \Sigma_{j} I\{d_m, d_j \in {\cal D}_v, & t(d_j)>t(d_m) \} \nonumber
\eqmoveup
\end{eqnarray}
where $I\{\cdot\}$ is the indicator function that takes a value of 1 when its
argument is true, and 0 otherwise. 
Intuitively,
$\zeta^{in}_m$ is the number of videos with meme $v$ that precede
video $d_m$ (potential sources), and $\zeta^{out}_m$ 
is the number of videos that succeed
meme $v$ after video $d_m$ (potential followers). 

The video influence
index $\chi_m$ is defined for each video document $d_m$
as the smoothed ratio of its out-degree over its in-degree,
aggregated over all meme subgraphs $G_v$ (Equation~\ref{eq:chii});
the smoothing factor 1 in the denominator accounts for $d_m$ itself).
The author influence index $\hat \chi_r$ is obtained by aggregating
$\chi_m$ over all videos from author $a_r$ (Equation~\ref{eq:chis}).
The normalized author influence index 
$\bar \chi_r$ is its un-normalized counterpart $\hat \chi_r$ divided by the number of
videos an author posted, which can be interpreted as the {\em average} influence of all
videos for this author.
\begin{eqnarray}\eqmoveup
\chi_m &=& \Sigma_{v} \frac{\zeta^{out}_{m, v}}{1+\zeta^{in}_{m,
v}} \label{eq:chii}\\
\hat \chi_r &=& \Sigma_{\{i, a(d_m)=a_r\}}~\chi_m, \label{eq:chis}\\
\bar \chi_r &=& \frac{\hat \chi_r}{ \Sigma_m I\{a(d_m)=a_r\} } 
\nonumber 
\end{eqnarray}

The influence indexes above captures two aspects in meme diffusion:
the volume of memes, and how ``early'' a video or an author is in the 
diffusion chain. The first aspect is similar to the {\em reweet} and 
{\em mention} measures recently reported on Twitter~\cite{cha2010measuring}.
The timing aspect in diffusion is designed to 
capture different roles that users play on Youtube,
These roles can be intuitively understood as 
 information {\em connectors} and {\em mavens}~\cite{gladwell2000tipping}.
Here {\em connectors} refer to people who come ``\ldots  
with a special gift for bringing the world together, 
\ldots [an] ability to span many different worlds'', 
and {\em mavens} are ``people we rely upon to connect us with new information, 
\ldots [those who start] word-of-mouth epidemics''.

The meme video and author graphs are used to generate features that describe 
node centrality and meme diffusion history, which are in turn used to predict 
importance of visual memes in Sec~\ref{sec:gfeatures}.

\secmoveup
\section{Predicting meme importance}
\label{sec:gfeatures}
\textmoveup

One long-standing problem in social media is on predicting the popularity of social memes~\cite{gladwell2000tipping}. Studies on social meme adoption and popularity has focused on URLs~\cite{Bakshy2011everyone}, hashtags~\cite{Yang2012hashtag}, or view counts on YouTube~\cite{Borghol2012clones,szabo2010popularity}. This work investigates whether or not visual meme popularity is predictable with knowledge of both the network and content.

Popularity, or importance on social media is inherently multi-dimensional, due to the rich interaction and information diffusion modes. For YouTube, it can be the number of times that a video is viewed~\cite{szabo2010popularity}, the number of likes or favorites that a video has received. While these commonly-used metrics focus on the entire video, not a given meme, we focus on two targets that are inherent to visual memes: the number of times that a video meme is reposted by other YouTube users (denoted as {\em volume}), or by the lifespan (in days) of a meme ({\em life}).  

We build a meme prediction model using features related to its early dynamics, the network around its authors, and the visual and textual content. 
For each visual meme $v$ that first appeared at time $t(v)$ (called {\em onset} time), 
we compute features on the meme- and author- sub-graphs up to 
$t_1 = t(v)+\Delta t$, by including video nodes that appeared before $t_1$. 
$\Delta t$ is set to one day in this work, to capture meme early dynamics, 
similar to what has been used for view-count prediction~\cite{szabo2010popularity}.

There are five types of features in total. For features aggregated over multiple authors, we take the maximum, average, median, and standard deviation among the group of authors who have posted or reposted the meme by $t_1$. (1) {\em volume-d1}, 1 dimension -- the volume of memes up to $t_1$.
(2) {\em connectivity}, 28 dimensions -- static network features of author productivity and connectivity.
We use the {\em total number of videos} 
that the author has uploaded to capture author productivity.
An author's connectivity include three metrics computed over the author and video graphs, respectively of up to time $t_1$: {\em degree
centrality} is the fraction of other nodes that a node is directly connected to; {\em closeness centrality} is the inverse of average path
length to all other nodes; and {\em betweenness centrality} is the fraction of
all-pairs shortest paths that pass through a node~\cite{brandes2001faster}. 
(3) {\em influence}, 16 dimensions -- dynamic features of author diffusion influence. These include the meme influence indices $\hat \chi_r$ and $\bar \chi_r$ in Equation~\ref{eq:chis}, 
as well as the aggregated in-degree and out-degree for each author. 
(4) {\em txt}, 2000 dimensions -- 
the bag-of-word vector for each meme $v$ is the average count of each word over all videos containing $v$ within the first day ; {\em vmeme} -- bag of visual meme vectors compiled in the same way as {\em txt}, with 2000 dimensions for the {\em Iran3} and 400 dimensions for {\em Swineflu}, respectively; {\em topic}, 50-dimensional vector is the posterior probability of each topic given $v$ inferred through the topic model in Section~\ref{sec:cm2}. 

We use Support Vector Regression (SVR)~\cite{libsvm} 
to predict meme volume and lifespan on a log-scale, using 
each, and the combination the features above. 


%% file: experiments.tex

\secmoveup
\section{Experiments and Observations}
\label{sec:exp}
\textmoveup

This section will first present our experimental setup, and then discuss our main results as both quantitative observations and as quantitative evaluations. The former include observations on video popularity versus remix probability (Section~\ref{ssec:repost}), and on apparent temporal diffusion patterns (Section~\ref{ssec:diffuidx}); the latter include visual meme detection (Section~\ref{ssec:neardups}), cross-modal annotation (Section~\ref{ssec:cm2result}), and popularity prediction (Section~\ref{ssec:predict}).

Using the targeted-querying and collection procedures described in
Section~\ref{sec:ytmonitor}, 
we downloaded video entries from about a few dozen topics from May 2009 to March 2010. 
We used the following four sets for evaluation, which had enough volume and change over time 
to report results, summarized in Table~\ref{tab:dataset}.  
The SwineFlu set is about the H1N1 flu epidemic. 
The Iran3 set is about Iranian domestic politics and related international
events during the 3-month period of summer 2009.  
The Housing set is about the housing market in the
2008-09 economic crisis; a subset of these videos were manually annotated and used 
to validate and tune the visual meme detection algorithms. 

We perform visual meme detection as described in
Section~\ref{sec:neardup}. We additionally filter the meme clusters identified by the 
detection system, by removing singletons
belonging to a single video or a single author.
We process words in the title and description of each video by 
morphological normalization with a dictionary~\cite{engdict}, 
we preserve all out-of-vocabulary words, these include foreign words and proper names 
(e.g., {\em mousavi}),  abbreviations ({\em H1N1}), or mis-spellings.
We rank the words by tf-idf across all topics, and take the top few thousand 
for topic models, tagging, and popularity prediction. 
The prototype system is implemented in C++, Python, and MATLAB, and it 
can be deployed on one workstation requiring less than 8GB memory.

\begin{table}[tb]
\begin{center}
\begin{tabular}{|@{\hspace{1mm}}l@{\hspace{1mm}}|@{\hspace{1mm}}c@{\hspace{1mm}}|@{\hspace{1mm}}c@{\hspace{1mm}}|@{\hspace{1mm}}c@{\hspace{1mm}}|@{\hspace{1mm}}c@{\hspace{1mm}}|}\hline
\small
 Topic        & \#Videos    & \#Authors   &  \#Shots       & Upload time  \\\hline
 SwineFlu    & 31,488    & 10,804    &  1,202,479   & 04/09$\sim$03/10 \\\hline
 Iran3         & 23,049    & 4,681     &  1,255,062   & 08/07$\sim$08/09 \\\hline
 Housing      & 2,446     & 654       &  71,872      & 08/07$\sim$08/09 \\\hline
\end{tabular}
\end{center}
\caption{Summary of YouTube event data sets.} \label{tab:dataset}
\captionmoveup
\end{table}

\begin{figure}[tb]
\centering \includegraphics[angle=0,width=.4\textwidth]{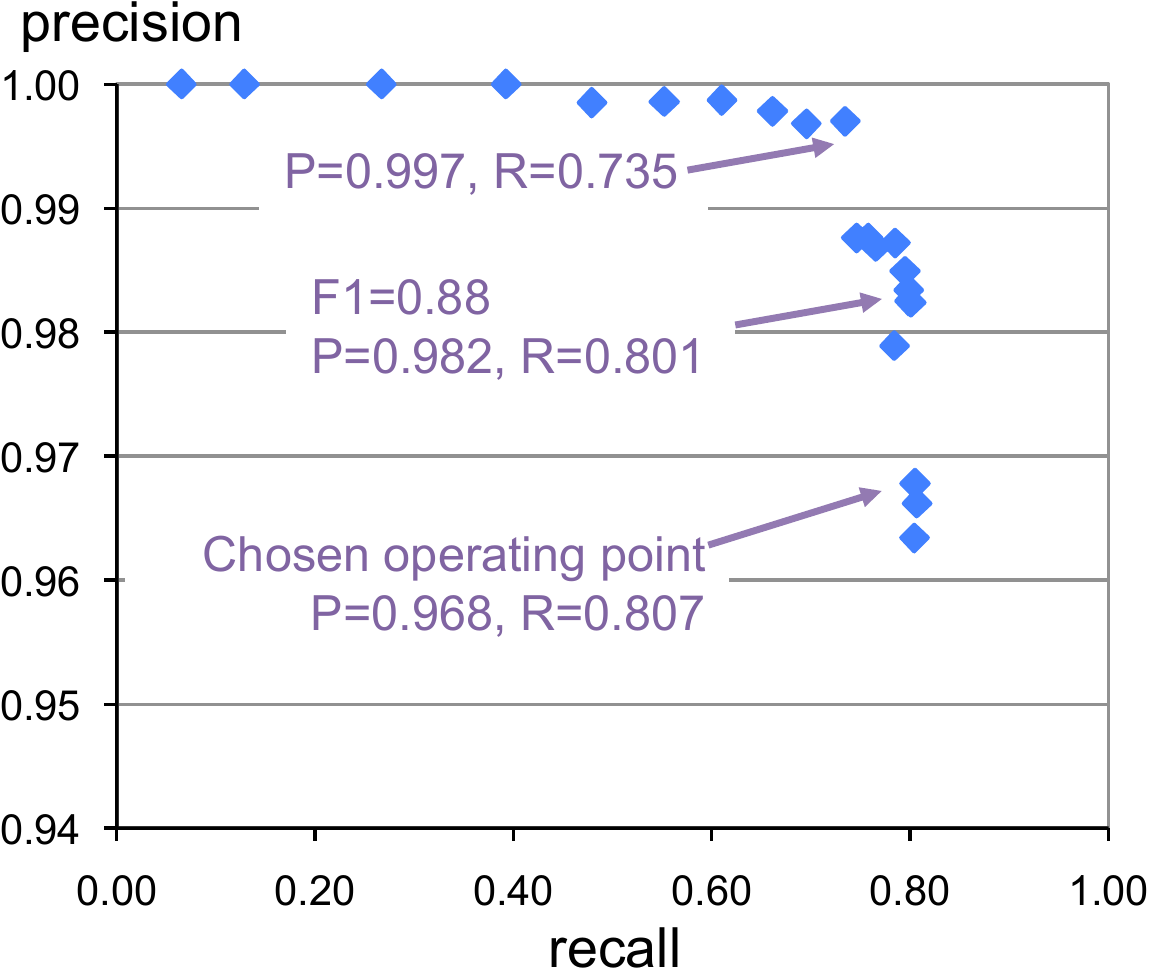} 
\captionmoveup
\caption{Performance of visual meme detection method on Housing Dataset.}
\label{fig:neardup} \captionmoveup
\end{figure}

\begin{figure*}[tb]
\begin{minipage}[b]{0.6\linewidth}
\centering \includegraphics[angle=0,width=\textwidth]{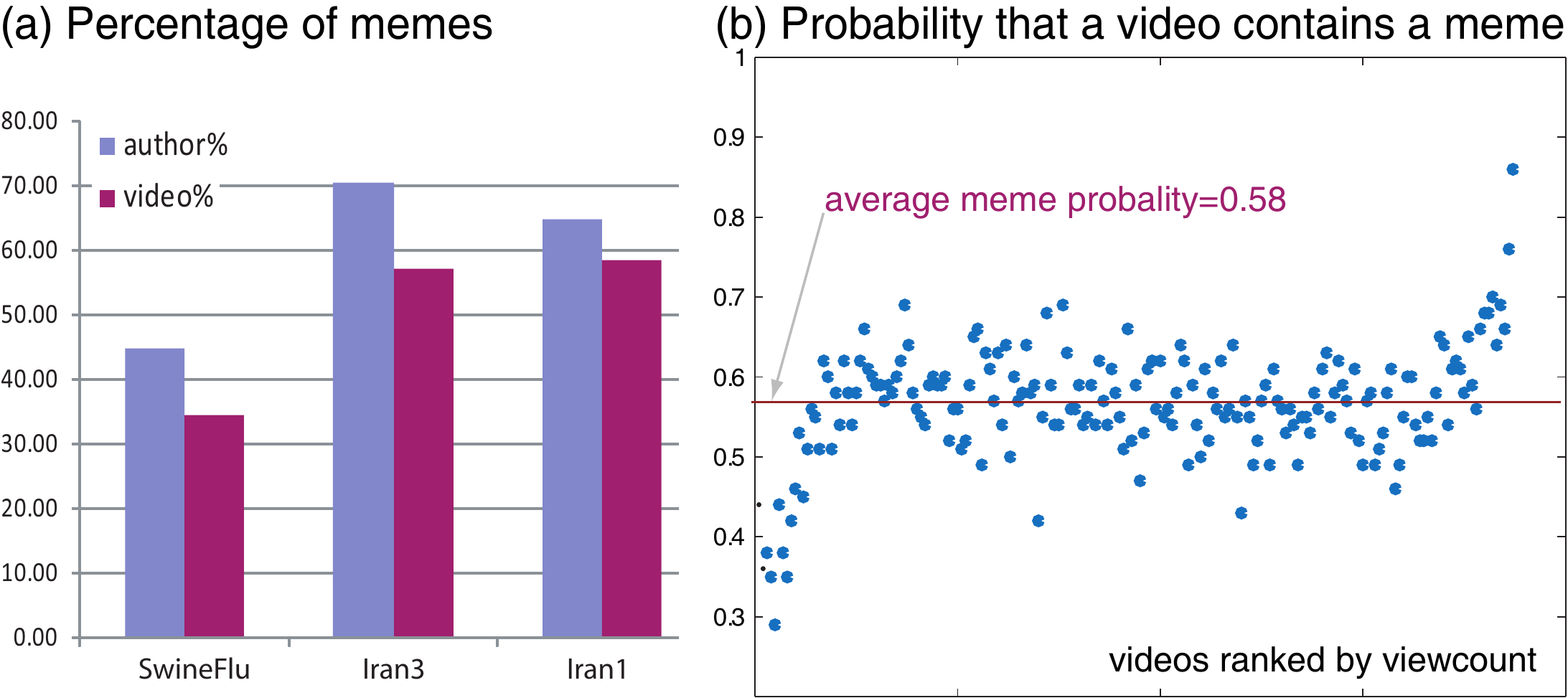}
\caption{Video reposting probabilities. (a) Fraction of visual memes. 
(b) Video views vs. meme probability on Iran3 set. }
\label{fig:repost} \captionmoveup
\end{minipage}
\hspace{2mm}
\begin{minipage}[b]{0.38\linewidth}
\centering \includegraphics[angle=0,width=\textwidth]{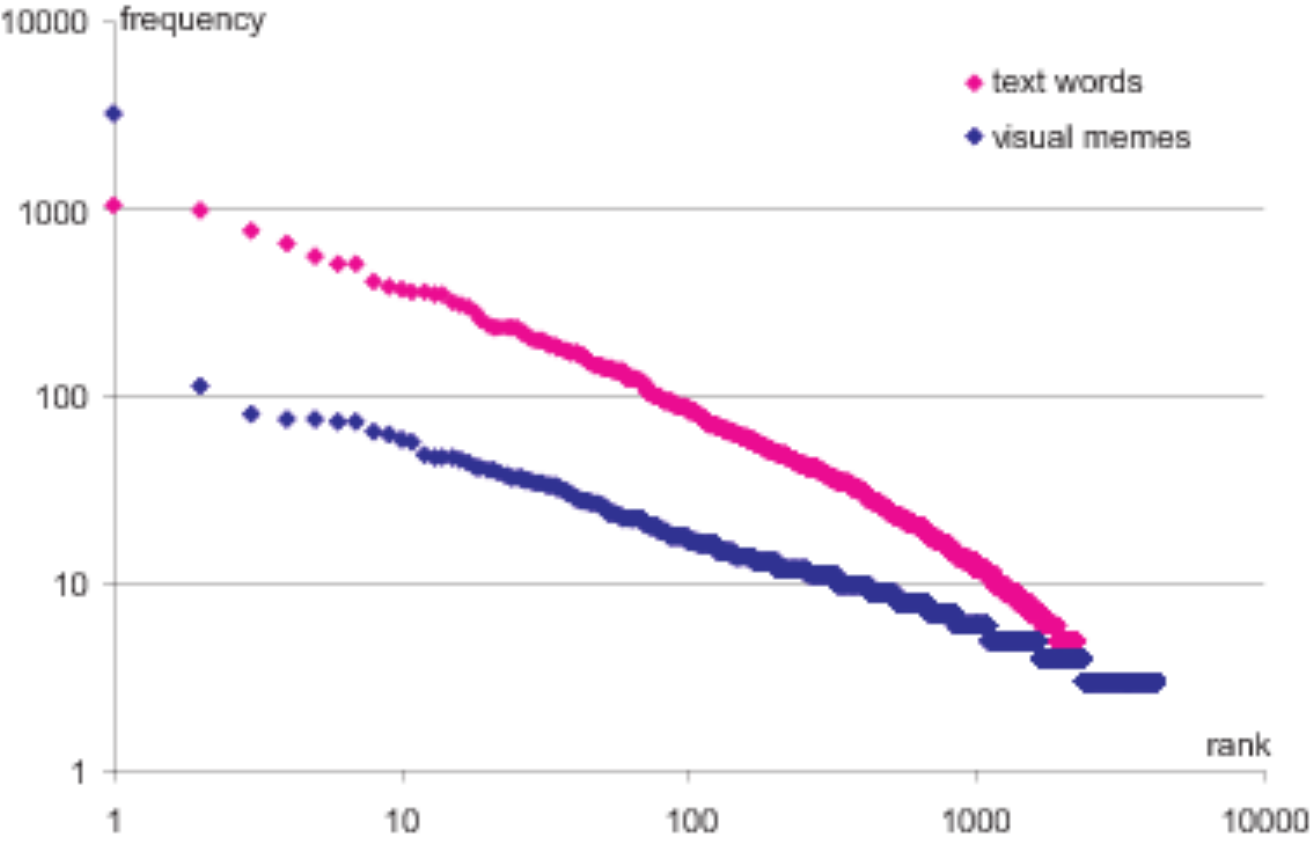}
\caption{Rank vs frequency for words and visual memes.}
\label{fig:wfreq}
 \end{minipage} \captionmoveup
\end{figure*}

\secmoveup
\subsection{Meme detection performance}
\label{ssec:neardups}
\textmoveup

We evaluate the visual meme detection method described in
Section~\ref{sec:neardup} using a test set created from the Housing dataset.
This is done by one annotator 
who is not aware of the detection algorithm, 
and she was instructed to find visually {\em very similar} keyframes
that are likely to be taken at the same scene. 
Specifically, she start from seed sets created from
multiple runs of k-means clustering with a tight cluster radius threshold, 
and top 50 returns based on color feature similarity using multiple random keyframes as the query. 
The annotator manually goes
through these results to explicitly mark the clusters as correct
and incorrect near-duplicates, and the top returns as duplicates with the query or not. 
This annotation protocol specifically includes many pairs 
that are being confused by the clustering and 
feature-similarity retrieval steps.
The resulting data set contains $\sim15,000$ near-duplicate keyframe pairs and
$\sim25,000$ non-duplicate keyframe pairs.

We compute the near-duplicate equivalence classes as described in
Section~\ref{sec:neardup}, and calculate
precision (P) and recall (R) on the labeled pairs. The results are shown on
Figure~\ref{fig:neardup} for varying values of the threshold
parameter~$\tau$.  We note that the performance is generally quite
high with $P>95\%$. There are several possible operating points,
such as $P=99.7\%,~R=73.5\%$ for low false alarm;
or $P=98.2\%,~R=80.1\%$ that produces
the maximum F1 score of $0.88$ (defined as $\frac{2PR}{P+R}$);
or $P=96.6\%,~R=80.7\%$ for the highest recall.
For the rest of our analysis, we use the last, high-recall, point with $\tau=11.5$.
On the Iran3 set of over 1 million shots, feature extraction takes around 7 hours 
on a quad-core CPU, and indexing and querying with FLANN takes 5 to 6 CPU hours.

\secmoveup
\subsection{Content views and re-posting probability}
\label{ssec:repost}
\textmoveup

In our video collections, the behavior of 
remixing and reposting is quite dominant.
Over 58\% of the videos collected for the Iran3 topic contain at least one visual meme,
and 70\% of the authors participate in remixing; 
likewise, 32\% and 45\% respectively for SwineFlu, as shown in Figure~\ref{fig:repost}(a). 
These statistics suggest that, for content covering trending events, 
a significant fraction of the authors 
re-mix and reprise existing sources.

Figure~\ref{fig:repost} 
shows empirical estimates of a video containing at least one meme in the Iran3 set,
binned by video view count (descending from left to right). 
We observe that the 4 most viewed videos have no memes and have nothing
to do with the topic, and likewise for 7 of the first 10.  One
has to get beyond the first 1,600 most popular videos 
before the likelihood of having
near-duplicates passes the average for the dataset,
at about $0.58$ (see Figure~\ref{fig:repost}(b)). 
There are several reasons for this mismatch. 
Among the video entries returned by YouTube search engine, 
the most viewed are often not related to the topic; 
and the search relevance criteria can be very different from 
factors that correlate with remix interactions. 
for example, the one with the highest view-count 
is a music video irrelevant to Iranian politics. 
In short, viewership popularity is a poor proxy for relevance or importance. 

\eat{
We have noted that this is a further example of the very unequal
distribution of views that characterizes this domain.  To
quantify the inequality of views-counts, we have
computed the Gini coefficient~\cite{gini1921measurement} of this data
set, and find it to have the extreme value of $0.94$, whether one looks at those
videos with near-duplicates, or those without.  The Gini coefficient,
used in economics, ranges from 0 (each video with an equal number of views)
to 1 (one video with all of the views).  The value we
observed far exceeds the measure of inequality for the distribution of
wealth in any known country, which has its maximum at about $0.7$.
}


We observe considerable similarity in the
frequency distribution of visual memes and words. 
Figure~\ref{fig:wfreq} is a scatter plot of textual words
and visual memes ranked by descending frequency in log-log scale.
Performing a regression fit, we obtain the following Zipf's
power law distributions:
$$f(w_t) \propto r^{1.102};~f(w_v) \propto r^{1.959} $$
The exponent $s$ for words in the title and description is close to
that of English words ($\sim1.0$). For visual memes \mbox{$s=1.959$},
suggesting that the diversity of visual memes is less than that of
words at the lower-frequency end.  
This suggest that it makes sense to model visual words 
appearances in a 
similar way as those with textual words.

\secmoveup
\subsection{Multimodal topics and CM$^2$}
\label{ssec:cm2result}
\textmoveup

We learn topic models on a joint vocabulary of words and memes.  For words, 
we adopt a tf-idf re-weighting scheme~\cite{manning2008introduction} across more than two dozen topics monitored around the same time, this is to suppress very popular words and yet not overly favor rare words. The visual meme vocabulary is constructed using a threshold on its frequency.
In the following experiments, we choose 12000 visual memes for the Iran3 and 4000 visual memes for SwineFlu collection, and 2000 text words for both collections. 

\begin{figure}[tb]
\centering \includegraphics[angle=0, width=0.5\textwidth]{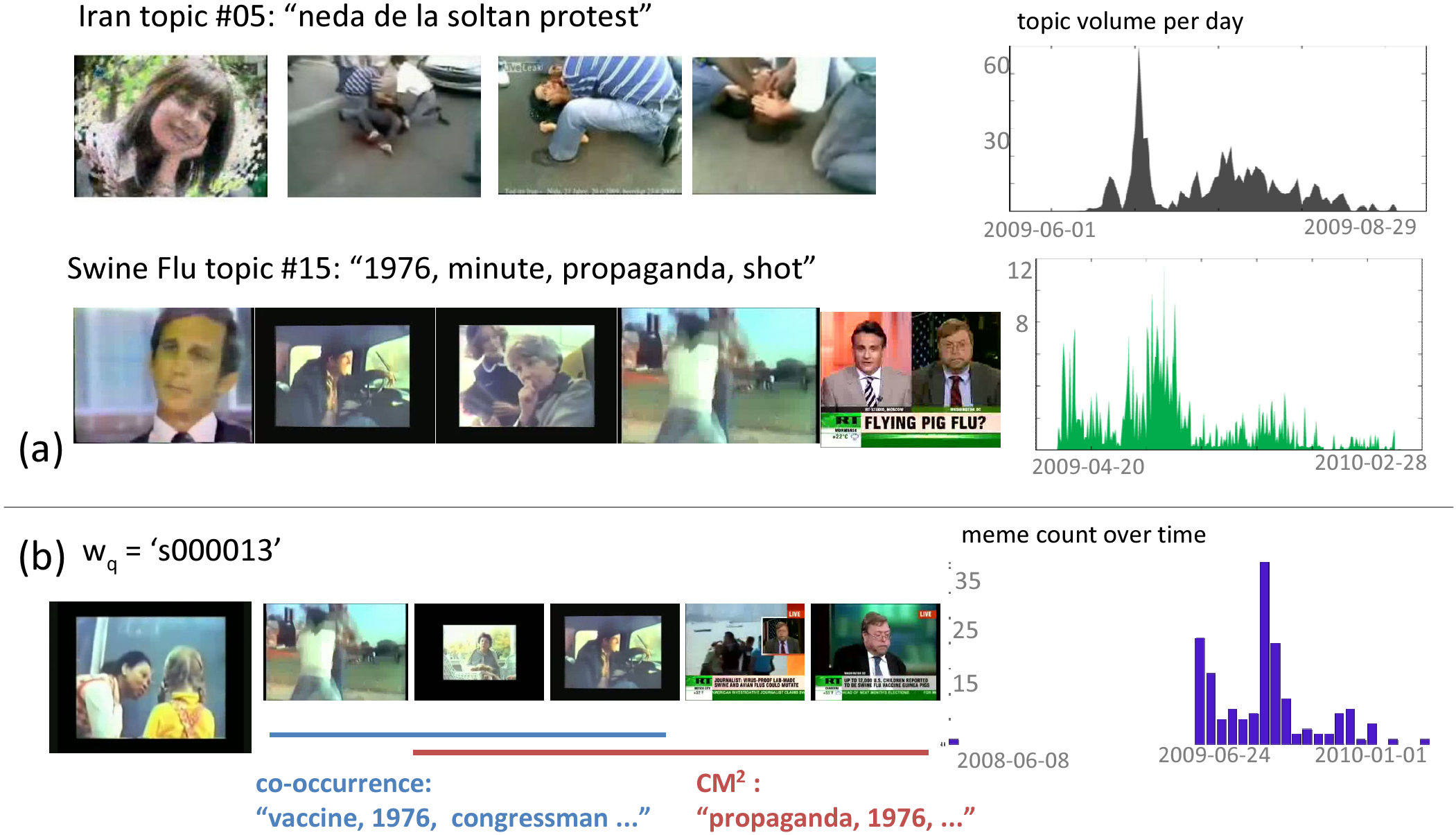}
\caption{(a) Topic model example. (b) Retrieval results using CM$2$. Please view in color and magnified for optimal readability.}
\label{fig:topic-cm2} \captionmoveup
\end{figure}

We set the number of topics to be 50 by validation, and use the term-topic probabilities $p(w|z)$ to label a topic, using both text words and visual memes. 
Figure~\ref{fig:topic-cm2}(a) shows two example topics over the collections Iran3 and Swineflu, respectively. Topic \#5 contains the keywords ``neda,
soltan, protest,\ldots'', the images capturing her tragic murder and her portrait that was published later. The topic volume over time
clearly showed the onset and peak of the event (just after June 20th, 2009), and we 
verified from the narrative on Wikipedia that this event also influenced subsequent protest activities in July, corresponding to another peak in meme volume.

We examine the CM$2$ model for video tagging in context. 
Here we consider using the visual memes of each video as the query 
and retrieve the tagging words using scores computed with Equation~\ref{eq:topic_tag}. We also implement the baseline in 
Equation~\ref{eq:word_tag} and look at the memes in comparison with those retrieved by top co-occurrence. We carried out 
five-fold cross-validation, and report the average performance based on the average log likelihood~\cite{barnard2003matching} of the existing tags. 
We did not use a precision or ranking metric, as tags associated with each video are sparse and positive-only, and many informative tags are missing in the data. 
Table~\ref{tab:tagperf} shows that the average log likelihood is significantly improved on both datasets, this demonstrates the advantage of the topic-based representation. 

\begin{table}[tb]
\begin{center}
\caption{Comparison on the log-probability of tags.} \label{tab:tagperf}
\begin{tabular}{ccc}\hline
\small
 Dataset        & Co-occurrence    		& CM$^2$   \\\hline \hline
 Iran3          & -6.08 $\pm$ 0.06    & -5.65 $\pm$ 0.05     \\\hline
 SwineFlu       & -6.59 $\pm$ 0.03    & -6.54 $\pm$ 0.04      \\\hline
\end{tabular}
\end{center}
\captionmoveup
\end{table}

Figure~\ref{fig:topic-cm2}(b) shows example retrieval result of using one of the
memes in the 1976 video as query. We can see that the co-occurrence
model returns {\em 1976, vaccine, congressman, \ldots} which are all
spoken or used as description in the original 1976 government propaganda video, 
while CM$2$ returns {\em 1976, propaganda}, which was apparently from the topic
above.
Comparing the images, we can also see that the top memes returned by
the co-occurrence model are all from the same video, since the parody
is mostly posted by itself, with little or no remix, while CM$2$
also retrieves two memes relating to modern-day vaccine discussions in
the news media, providing relevant context.

The rightmost column in Figure~\ref{fig:topic-cm2} shows the temporal evolution of a meme (Figure~\ref{fig:topic-cm2}(b)) and two topics (Figure~\ref{fig:topic-cm2}(a) ). We
can see the source of the 2008 propaganda video in the meme evolution, 
it also reveals that there are multiple waves of remix and
re-posting around the same theme. The topic evolution, 
on the other hand, segments out sub-events from the broader unfolding of many
themes -- with Iran topic \#5 representing the murder of Neda and its subsequent influence, and Swine Flu \#15 closely correlated to the
public discussion on vaccines.

\secmoveup
\subsection{Influence index of meme authors}
\textmoveup
\label{ssec:diffuidx}

\begin{figure}[tb]
  \centering \includegraphics[angle=0,width=.48\textwidth]{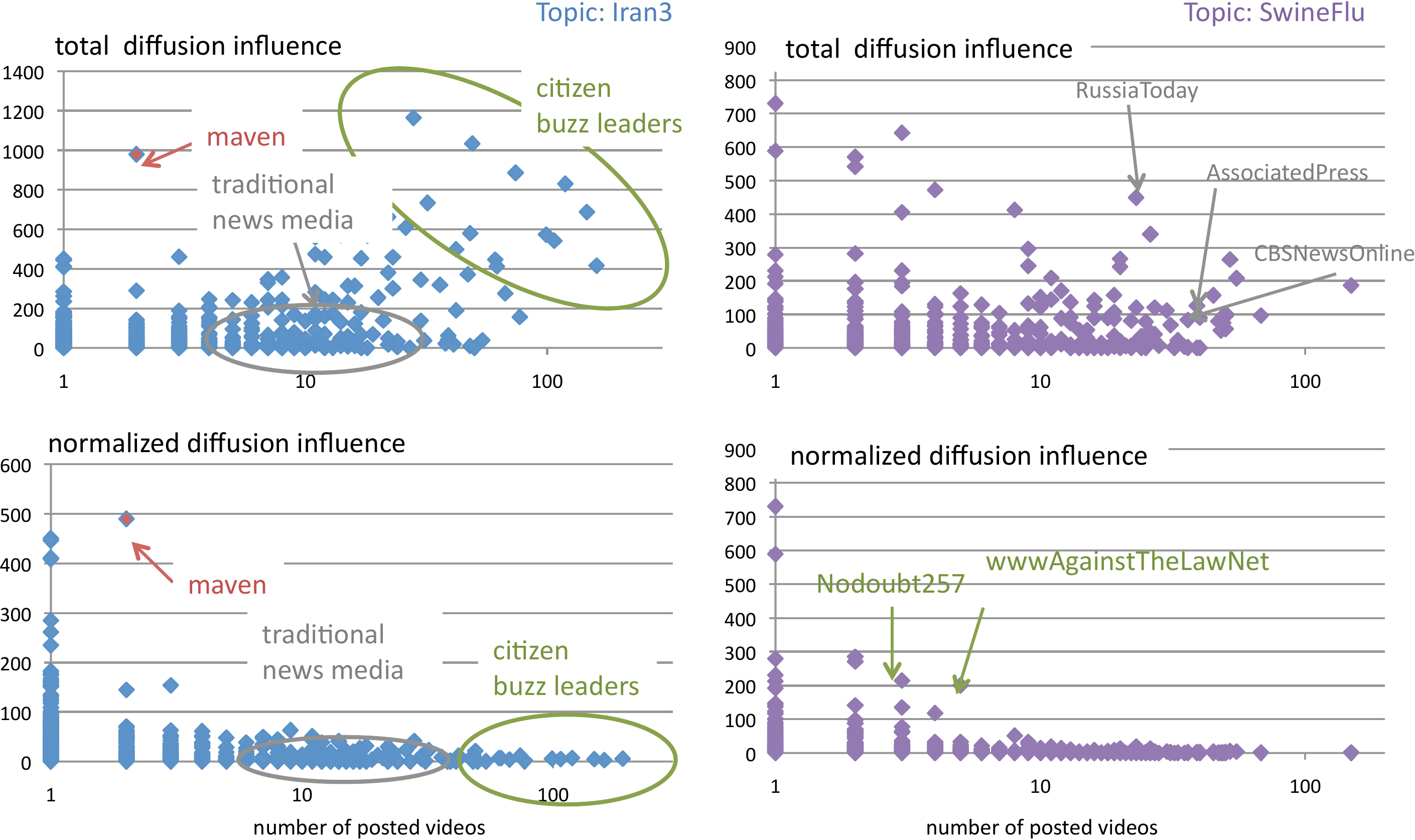}
  \caption{Meme influence indices vs author productivity on 
  topic Iran3 (Left) and SwineFlu (Right); detailed discussions in Sec~\ref{ssec:diffuidx}. We recommend viewing in color and with magnification.}
  \label{fig:diffusion} \captionmoveup
\end{figure}  

We compute the diffusion index for authors according to
Equation~\ref{eq:chis}. Figure~\ref{fig:diffusion} contains scatter plots 
of the author influence indices on the $y$-axis, versus 
``author productivity'', 
(number of videos produced) on the $x$-axis. 
For both the Iran3 topic and the SwineFlu topic, we plot
the total diffusion influence $\hat \chi_r$
and the normalized diffusion influence $\bar \chi_r$. 

In the Iran3 topic we can see two distinct types of
contributors.  We call the first contributor type {\em maven}~\cite{gladwell2000tipping} (marked in red), 
who are post only a few videos, which tend to be massively remixed and reposted. 
This particular maven was among the first to post the murder of Neda Soltan,
and one other instance of student murder on the street.  The
former post become the icon of the entire event timeline.  We call the second
contributor type information {\em connectors}~\cite{gladwell2000tipping}
(circled in green), who tend to produce a large number of videos, 
and who have high total influence factor, yet has lower influence per video. 
They aggregate notable content, and serve the role of bringing this
content to a broader audience.
We examined the YouTube channel pages for a few authors in this
group, and they seem to be voluntary political activists with
screennames like ``iranlover100''; and we can dubb them ``citizen buzz leaders''.  
Some of their videos are slide shows of iconic images. 
Note that traditional news media, 
such as AljezeeraEnglish, AssociatedPress, and so on (circled
in gray) have rather low influence metric for this topic, partially because the Iran
government severely limited international media participation in the
event; most of the event buzz was driven by citizens.

However, the SwineFlu collection behaves differently in its influence
index scatterplots. 
We can see a number of {\em connectors} on the upper right hand side of the total
diffusion scatter. But it turns out that they are the traditional media (a few marked in gray), most of which have a large number ($>$40) of videos with memes. 
The few {\em mavens} in this topic 
(marked with green text) 
are less active than in the Iran topic, and notably they all
reposted the identical old video containing government health propaganda 
for the previous outbreak of swine flu in 1976. 
These observations suggest that it is the traditional new media 
who seem to have driven most content on this topic, 
and that serendipitous discovery of novel content does also exist, 
but has less diversity. 

These visualizations can serve as a tool to characterize influential users 
in different events. We can find user groups serving as {\em mavens}, 
or early ``information specialists''~\cite{gladwell2000tipping}, and {\em connectors},
who ``brings the rest \ldots together'', and henceforth observe different 
information dissemination patterns in different events.

\begin{figure*}[tb!]
\centering
\begin{minipage}{.95\linewidth}
  \centering \includegraphics[angle=0,width=.97\textwidth]{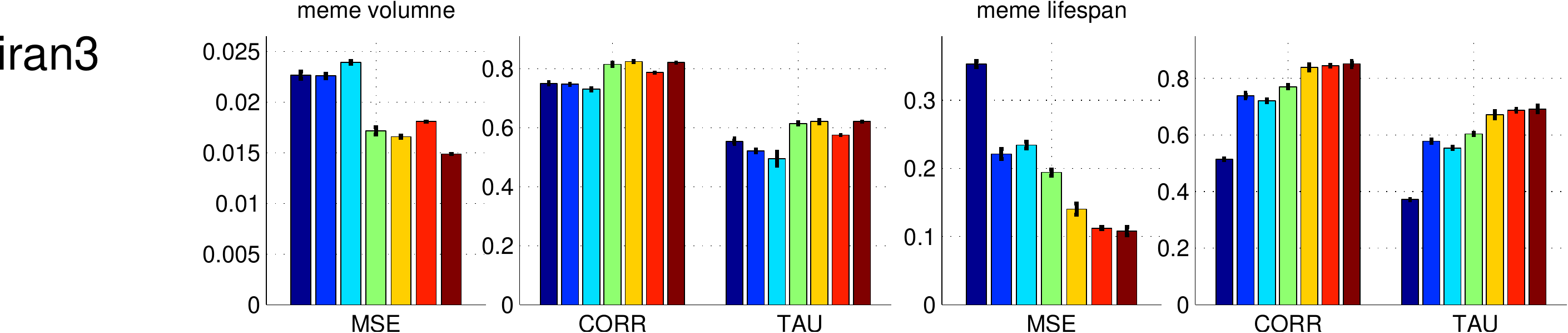}
\end{minipage}
\begin{minipage}{.95\linewidth}
  \centering \includegraphics[angle=0,width=\textwidth]{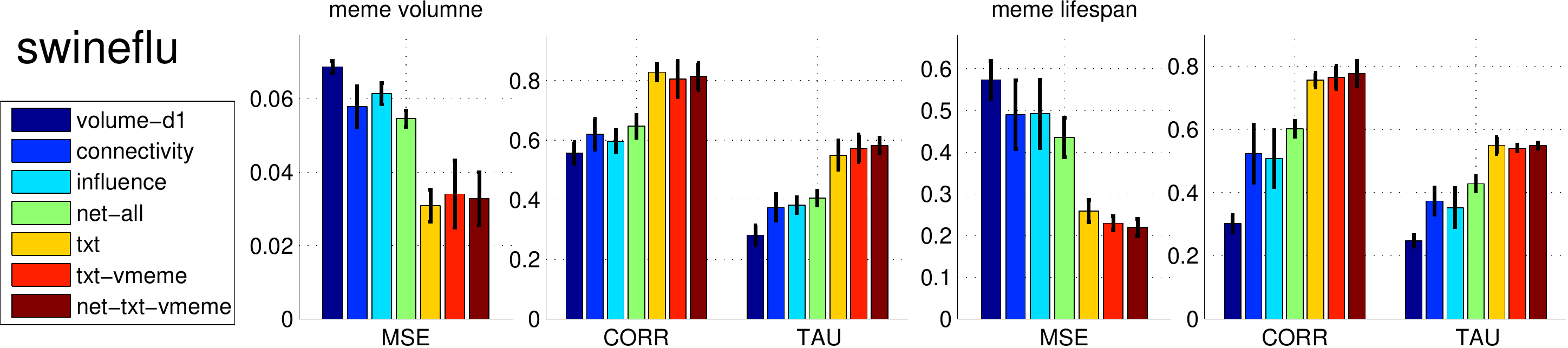}
\end{minipage}
\caption{Meme popularity prediction performance using various network and content features (best viewed in color). Top: Iran3 dataset; bottom: Swineflu dataset. Prediction targets: meme volume (\# of videos, left) and lifespan (days, right); performance metrics: M.SE (smaller is better) pearson correlation and Kendall's tau (larger is better). See Section~\ref{ssec:predict} for discussions on various features.}
\label{fig:meme_pred} \captionmoveup
\end{figure*}

\secmoveup
\subsection{Meme popularity prediction results}
\label{ssec:predict}
\textmoveup

We predict the lifespan of memes as described in Section~\ref{sec:gfeatures}. 
We prune memes that appear less than 4 times, the remaining memes are randomly split in half for training/testing. The Iran3 dataset has 4081 memes in each of the train/test split, the SwineFlu set has 398. Different features are filtered by a low correlation threshold (0.03) and then concatenated to form a joint feature space. We train support vector regressor~\cite{libsvm} by searching over hyperparameters C and several kernel types -- linear, polynomial, and radial basis function of different width. We use three different metrics to quantify regression performance: mean-square-error ({\tt mse}), Pearson correlation ({\tt corr}), Kendall's tau ({\tt tau})~\cite{kendall1938new}. Each regressor is trained with five different random splits of train/test data, the average performance with their standard deviation (as error wisks) is shown in Figure~\ref{fig:meme_pred}. 

\begin{figure*}[tb!]
\begin{minipage}{.48\linewidth}
  \centering \includegraphics[angle=0,width=\textwidth]{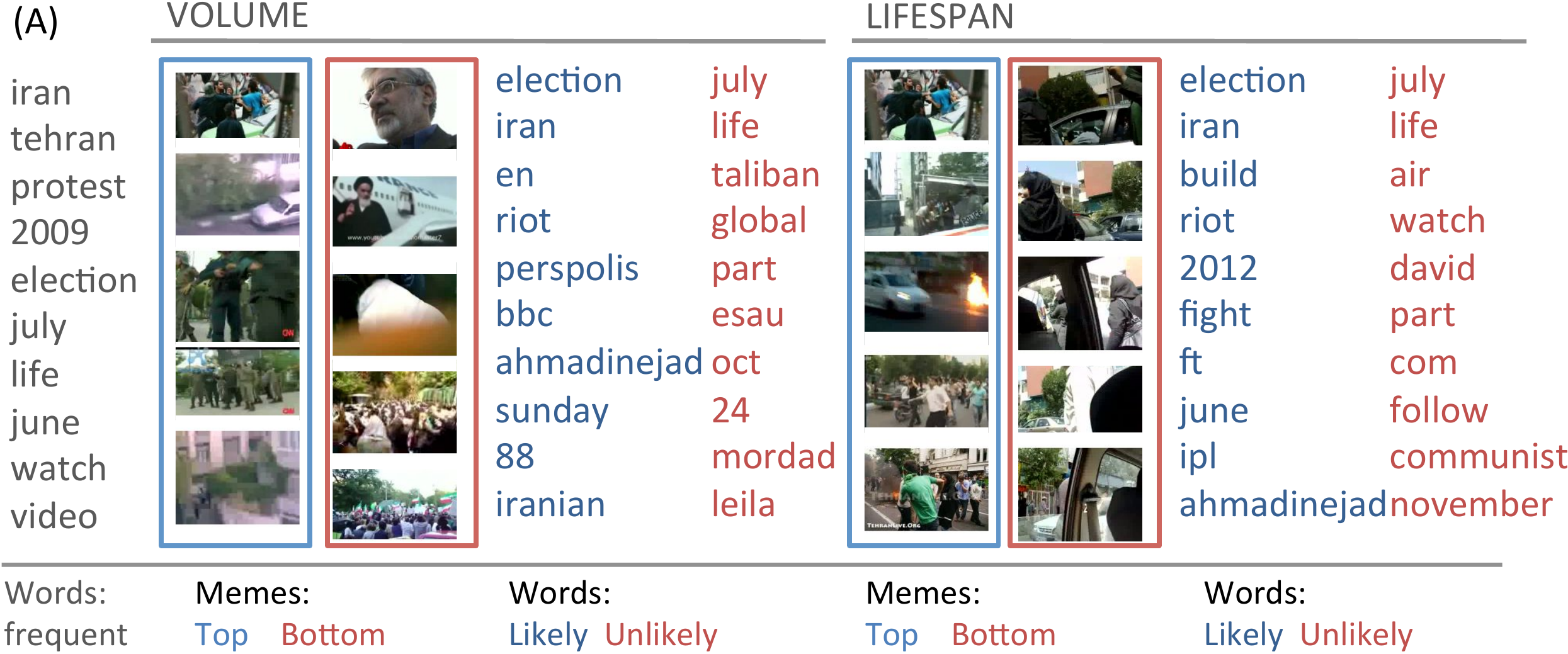}
\end{minipage}
\begin{minipage}{.48\linewidth}
  \centering \includegraphics[angle=0,width=\textwidth]{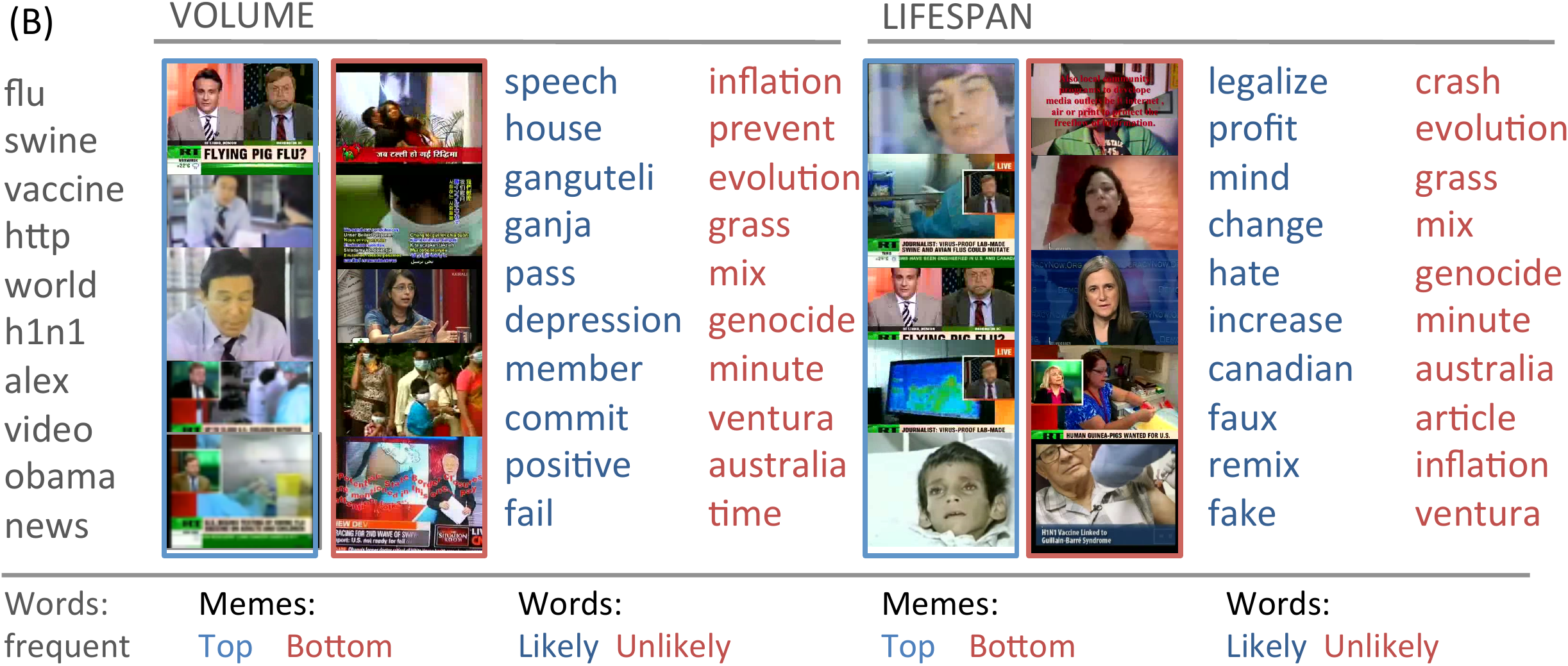}
\end{minipage}
\caption{Illustration of meme popularity prediction results (best viewed in color). (A) Iran3 dataset; (B) SwineFlu dataset. Top popular (blue) and unpopular (red) memes by prediction score; and most likely (blue) and unlikely (red) words by correlation. } \captionmoveup
\label{fig:meme_expl}
\end{figure*}

We can see that 
meme graph features ({\em connectivity} and {\em influence}) both out-perform the 
baseline feature {\em volume-d1}. 
Note that {\em volume-d1} is the conceptual equivalent of the
early view-count features that Szabo and Huberman~\cite{szabo2010popularity}
used to predict long-term view-count on Youtube.
Combining these three types of features ({\em net-all}) further improves prediction performance, and text keyword feature ({\em txt}) is single strongest predictor. The presence and absence of other visual memes is not stronger than text ({\em txt+vmeme}), while all of network, words and meme features has the strongest combined performance ({\em net+txt+vmeme}). The Iran3 dataset, with significantly more data to learn from, has better and more stable results than SwineFlu. From the average MSE, the predictions for meme volume on Iran3 is within 1.7\% ($\sqrt{ 10^{.015}}$) and 16.1\% ($\sqrt{ 10^{.13}}$) for meme lifespan. In Figure~\ref{fig:meme_expl} we examine the top- and bottom- ranked visual meme with {\em net+txt} feature, showing that the top memes are intuitively on-topic, while most of the low-ranked memes have no clear connection to the topic. Figure~\ref{fig:meme_expl} also shows the positively and negatively correlated words to each of the 
prediction target. We can see that they are notably different from frequently-occurring words in either collection. Indicative words include those that indicate trustable authors ({\em bbc}), particular sub-events ({\em riot}), or video genre that engender participation ({\em remix}). On the other hand, certain frequent words such as {\em watch, video}, or {\em swine flue, h1n1} are shown to be non-informative. 

%% file: related.tex
\bigsecmoveup
\section{Related Work}
\label{sec:related}
\textmoveup

\eat{
\TODO{from R2: ``multimodal tag prediction, predicting roles of users in social networks, sampling a representative data collection from a social network''}

\TODO{from R1 ``It should be structured to review in different paragraphs relevant works in the following areas: a) meme tracking in social media, b) social video analysis, c) near-duplicate videos detection and d) predicting influence in social media. In addition, how do the actual techniques used in the paper (Sec 3 and 4) compare to techniques in other works related to video analysis and near-duplicate video detection? Something is mentioned along the text in Sec 3 and 4, but should be better explained in Sec 2.''}

\TODO{from R3: The main problem of this paper, tracking and modeling information diffusion, has been a very important topic in many research communities. The paper omitted citations for the papers which studied very similar problems. For example, for predicting meme popularity or lifespan:}
}

This work is related to several active research areas in  
understanding social media and analyzing multimedia content.

Memes have been studied in online information networks of modality other than videos. 
Retweeting on micro-blogs is a common example~\cite{Kwak10twitter}, 
where users often quote the original text message verbatim, 
having little freedom for remixing
and context changes within the 140 character limit.
MemeTracker~\cite{leskovec2009meme}
detects quoted phrases among millions of blogs and news posts. 
Kennedy and Chang~\cite{Kennedy08imagecopy} 
detects edited images from web search results.
Several measurement studies tracked explicit social interactions around online videos. 
Cha et al.~\cite{cha07youtube}
characterized content category distribution and exact duplicates of popular videos.
Subsequent studies on YouTube include tracking
video response actions~\cite{benevenuto2009video} using metadata, 
and modeling user comments to determine
interesting conversations~\cite{DeChoudhury2009interesting},
The audiovisual content of online videos 
are used to analyze
individual social behavior such as personality~\cite{Biel_ICWSM2010},
or used to generate content summaries~\cite{hong11beyondsearch}.
Schmitz et al.~\cite{schmitz2006international} showed that the frequency remix 
for film content can be used as an implicit video quality indicator.
Detecting visual memes on a large-scale and 
using them to infer implicit social interaction has not been done before. 

Our method for detecting visual memes builds upon those for 
tracking near-duplicates in images and video. Recent foci in near-duplicate detection 
include 
speeding up detection on image sequence, frame, or local image
points~\cite{wu09neardupelimination}, exploring the effect of 
factors other than visual features~\cite{natsev2010design}, and  
scaling out to web-scale computations using large compute clusters~\cite{liu07clustering}.
Compared to copy detection, our work tracks mutual remixes in a large collection 
rather than matching one query video with reference database. 
Our method is similar to existing approaches in feature choice, 
and using approximate nearest-neighbor indexing enables scaling 
to more than 1 million shots. 
Topic models~\cite{barnard2003matching} 
is among the popular techniques for the joint modeling of images and text. 
Our new insight on the CM$^2$ models is that nearest-neighbor pooling on the topic space 
works better than direct inference on topic models, likely due 
to the noisy nature of social media text.

Social media popularity and user influence is a very active 
research area, and our study is a special cases on visual meme.
Ginsberg et al.~\cite{ginsberg2008detecting} showed that 
web query popularity is correlated with real-world trends such as flu. 
For text memes, Yang and Leskovec~\cite{yang2010modeling} 
proposes a linear model to predict meme volume, and further quantifies the temporal shape of meme evolution~\cite{yang2011patterns}.
Other factors that influence popularity include user and the underlying network~\cite{Yang2012hashtag}
the nature of the topic~\cite{Romero2011hashtag}, and sentiment associated with 
the message~\cite{Bakshy2011everyone}. 
For views on YouTube, Crane and Sornette~\cite{crane2008robust} characterize the driving mechanisms of 
YouTube views as driven by external events or internal means (virality). 
Szabo and Huberman~\cite{szabo2010popularity} used early video views 
to predict longer-term view counts.
Borghol et al.~\cite{Borghol2012clones} studied whole-clip video clones 
to quantify content-agnostic factors that influence video views. 
For describing user roles in information diffusion, 
Basky et al.~\cite{bakshy2009social} describes early adopters and influencers for
spreading user-created content in virtual communities,
Saez-Trumper et al~\cite{saez2012finding} defines trend-setter using graph metrics.
Part of our contribution is in demonstrating evidence that fine-grained content features
are effective for predicting popularity at individual message level (within a topic).


%% file: conclusion.tex

\secmoveup
\section{Conclusions}
\label{sec:conclusion}
\textmoveup

In this paper, we proposed visual memes for tracking and
monitoring of real-world events on YouTube. We described a
system for monitoring event traces in social video, 
and proposed a scalable algorithm for extracting visual memes with 
high accuracy. 
Our system shows that detecting visual memes at a large scale
is feasible. Furthermore, we have observed significant remix-behavior 
in videos discussing news events (up to 70\% authors and 60\%),
and observed different users roles 
in propagating information.
We designed a cross-modal matching (CM$^2$) method 
for annotating the meanings of visual memes, and 
have observed that social network and content features both contribute to 
better predictions of meme popularity. 
We release an open-source 
implementation of the meme-detection algorithm~\cite{projpage}.
While the original video data are not available due to 
copyright and YouTube terms of service, 
we have made the list of YouTube video ids available, 
along with the meme detection results~\cite{projpage}. 

We would like to point out some limitations of this work.
Patterns on two long-running news topics are intended as case studies, 
there is more work needed before the conclusions generalizes 
to all events or online video-making in general. The scope and 
operational definition of video memes in this work is limited to 
video remixing that largely preserves the visual appearance 
and semantic content. One can potentially consider other definition 
of visual memes with larger variations in both appearance and semantics, 
such as ``Dancing Matt''\footnote{\surl{http://en.wikipedia.org/wiki/Matt_Harding}}, 
a YouTube phenomena with an iconic person and action performed in different locations.
The quantitative evaluation of video meme detection is base on
annotation from one trained annotator, the reliability of the annotation
task is unknown as presented in this work.

Future work can take several directions. 
For example, we may leverage visual memes for better annotations 
and content search for online videos. 
We may also examine the sequence and timing of meme shots in relation to popularity and likelihood of a remix. Finally, we are interested in examining visual remixes in video genres other than news. 